# Hysteretic phenomena in GFET: general theory and experiment

Anatolii I. Kurchak[1], Anna N. Morozovska[2], and Maksym V. Strikha[1,3]*

[1] *V.Lashkariov Institute of Semiconductor Physics, National Academy of Sciences of Ukraine,*
*pr. Nauky 41, 03028 Kyiv, Ukraine*
[2] *Institute of Physics, National Academy of Sciences of Ukraine,*
*pr. Nauky 46, 03028 Kyiv, Ukraine*
[3] *Taras Shevchenko Kyiv National University, Radiophysical Faculty*
*pr. Akademika Hlushkova 4g, 03022 Kyiv, Ukraine*

We propose a general theory for the analytical description of versatile hysteretic phenomena in a graphene field effect transistor (GFET) allowing for the existence of the external dipoles on graphene free surface and the localized states at the graphene-surface interface. We demonstrated that the absorbed dipole molecules (e.g. dissociated or highly polarized water molecules) can cause hysteretic form of carrier concentration as a function of gate voltage and corresponding dependence of graphene conductivity in GFET on the substrate of different types, including the most common $SiO_2$ and ferroelectric ones. It was shown that the increase of the gate voltage sweeping rate leads to the complete vanishing of hysteresis for GFET on $SiO_2$ substrate, as well as for GFET on ferroelectric substrate for applied electric fields *E* less than the critical value $E_c$. For $E>E_c$ the cross-over from the hysteresis to anti-hysteresis takes place. These results well correlate with the available experimental data up to the quantitative agreement. Proposed model takes into consideration the carriers trapping from the graphene channel by the interface states and describes the "anti-hysteresis" in GFET on PZT substrate well enough. Obtained results clarify the fundamental principles of GFET operation as well as can be directly applied to describe the basic characteristics of advanced non-volatile ultra-fast memory devices using GFET on versatile substrates.

---

* corresponding author, e-mail: maksym_strikha@hotmail.com

Linear spectrum of graphene causes remarkable symmetry in dependence of its conductivity on gate voltage: $\sigma(V_g) = \sigma(-V_g)$ [1, 2, 3]. However, this symmetry occurs in the highest quality exfoliated graphene with specially prepared surface and interface with substrate. For less perfect graphene, e.g. for the one produced by CVD and used for applications in devices, many studies (see e.g. refs. [4, 5, 6, 7, 8, 9, 10, 11, 12, 13, 14, 15, 16, 17, 18, 19, 20, 21, 22, 23]) reports about the hysteresis dependence of graphene conductivity on the gate voltage. The direction of the hysteresis loop could be different, namely it can be a direct or an inverse ones (anti-hysteresis). The type of hysteresis and the form of the loop depend on the sweeping rate of the switching gate $V_g$, defined as the derivative $dV_g/dt$, as well as on the gate voltage range $V_g^{sw}$ and on the conditions at the graphene surface and interface.

The presence of hysteresis in the $V_g$-dependence of graphene channel conductivity enables the creation of two strongly different states with different resistivity of the channel, "0" and "1" for logical units. Understanding of the physical mechanism underlying in the hysteresis phenomena is crucial for design of the combined systems "graphene-on-surface" for advanced ultrafast non-volatile memory based on the graphene field effect transistor (GFET).

In the first works devoted to the problem [,] [15 - 18] hysteresis was found in the system "graphene channel on SiO$_2$ substrate". However, the system works at high switching voltage (~70V) and therefore cannot be used in real non-volatile memory units. However, the usage of different ferroelectrics (like Pb(Zr$_x$Ti$_{1-x}$)O$_3$ (PZT) [,], Ba$_{1-x}$Sr$_x$TiO$_3$ (BSTO) , etc.) with a dielectric permittivity κ much greater than that of SiO$_2$ (κ = 3.9) allows to create an ultrafast non-volatile memory of new generation [,] [19] [15,]. However, the quantitative description of the new memories requires a profound understanding of several rival mechanisms physics, which create different type of hysteresis, direct or inverse one (or so-called "anti-hysteresis").

First of all, if the graphene channel is chemically doped by electrons or holes, the symmetric relation $\sigma(V_g) = \sigma(-V_g)$ violates. For instance, CVD fabrication of graphene that leads to the chemical doping of graphene channel by holes [,] , causes the shift of electro-neutrality point on $\sigma(V_g)$ from $V_g = 0$ to positive voltages.

Another reason for graphene doping can be polar molecules (e.g. water ones) absorbed on graphene surface [,] [15 - 18, 24, 25, 26, 27]. Different substances can cause different type of doping. In particular absorbed NH$_3$ and CO molecules dope the graphene channel with electrons, and hence shift of electro-neutrality point to negative voltages range, while H$_2$O and NO$_2$ dope it with holes and the positive shift of the electro-neutrality point appears.

The doping can be caused by the spontaneous polarization of ferroelectric substrate, at that the electro-neutrality point shift from $V_g = 0$ can be either right or left depending on the polarization direction. The characteristic hysteresis in the conductivity $\sigma(V_g)$ of graphene-on-ferroelectric is determined by the non-linear response of ferroelectric dipoles to external electric field, caused by the gate voltage ˒.

The last, but not the least factor to be taken into account is the interface structure between a graphene and a surface. Usually there can be a great number of interface states which can, depending on the ambient conditions, either trap the carriers from graphene, or inject them into it. The effect can cause anti-hysteresis in $\sigma(V_g)$ dependence [28, 29].

Despite the large number of studies on this subject, only a few theoretical models, which describes hysteresis in GFET and consider the presence of surface dipoles on the free surface of graphene , as well as interface states at the graphene-substrate ˒, have been proposed. However, ˒ to describe anti-hysteresis in GFET on ferroelectric PZT substrate a quite narrow range of gate voltages, $E<E_c$, was considered, and so the contribution of ferroelectric polarization hysteresis was not taken into account, but only the gate doping given the large dielecric permittivity κ for PZT.

To resume the review, despite the great amount of efforts focused on the problem, there is no comprehensive analytical model of hysteretic phenomena in GFETs taking into account all the abovementioned rival factors. The gap in the knowledge motivated us to evolve a general theory for the quantitative description of versatile hysteretic phenomena in GFETs on different substrates. The phenomenological approach presented below considers the existence of both the external dipoles on graphene free surface and the localized states at the graphene-surface interface. Our approach takes into account multiple factors, which make a significant contribution to the conductivity of the graphene channel on substrates of various types, namely: a) the presence of surface contaminants (dipoles) on the free surface of graphene and their dynamics depending on the rate of gate voltage switching and relaxation time of dipoles for GFET systems with ferroelectric and $SiO_2$ substrates; b) the presence of ferroelectric polarization hysteresis depending on the magnitude of applied electric field considered in the framework of nonlinear Landau-Khalatnikov relaxation equation; c) the presence of interface states in graphene-substrate interface.

**Theoretical formalism**

We study a heterostructure-type system consisting of a conducting graphene channel, placed on a dielectric or ferroelectric substrate (see **Figure 1**). The concentration of 2D carriers in the channel is governed by several factors, which are a time-dependent gate voltage $V_g(t)$, polarization of the dipoles of a ferroelectric substrate, polarization of dipoles (e.g. absorbed water molecules) on graphene surface, and the charge carriers localized by traps at graphene-substrate interface. Later we assume that the time dependence $V_g(t)$ is saw-like Eq. (A.1) with a period $2\pi/\omega$ and the amplitude $A$ (see Fig.A.1 in the Supplementary Materials [30]).

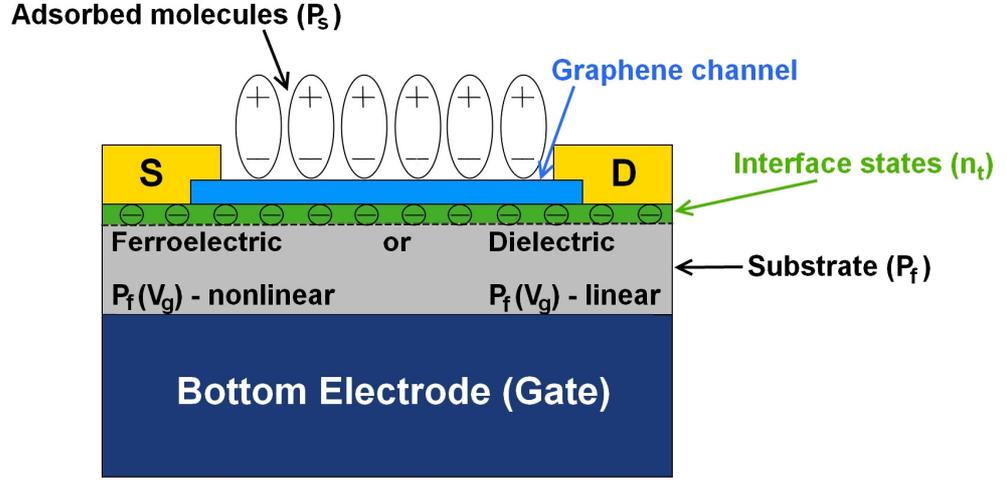

**Figure 1.** Considered system includes a graphene conducting channel placed on a dielectric or ferroelectric substrate. Dipoles (e.g. polar water molecules) can be absorbed by the graphene free surface. Carriers from the channel can be trapped by the centers localized at graphene-substrate interface.

Allowing for all the abovementioned factors, the resistivity of graphene channel can be presented similarly to :

$$\rho[V_g, P_s\ P_f, T] \approx \frac{1}{\sigma(V_g, P_s\ P_f, T)} + \frac{1}{\sigma_{intr}(T)} + \frac{1}{\sigma_{min}}, \qquad (1)$$

where $P_s(t)$ is the polarization of dipoles absorbed on graphene surface, $P_f(t, T)$ is the temperature-dependent ferroelectric polarization. In the most common case for graphene under ambient conditions the scattering of ionized impurities by substrate dominate [,31], and so the conductivity of 2D graphene channel can be presented as $\sigma(V_g, P_s\ P_f, T) = e\mu n(V_g, P_s\ P_f, T)$, where $n(V_g, P_s, P_f)$ is 2D carriers concentration per unit area, caused by the gate mixed doping, and by the dipoles absorbed by the surface, as well as by ferroelectric dipoles; $\mu$ is carriers mobility that is also a function of temperature $T$ in a general case. The second term in Eq.(1) corresponds the intrinsic graphene conductivity that should be taken into consideration, when Fermi level is in the vicinity of electro-neutrality point, $\sigma_{intr}(T) = e\mu n_{intr}(T)$, $n_{intr}(T) = \frac{2(k_B T)^2}{\pi(\hbar v_F)^2}$. The third term in Eq.(1) corresponds the minimal quantum conductivity, $\sigma_{\min} \approx \frac{4e^2}{\hbar}$, that should be taken into consideration at low *T*. Fermi level position in a single-layer graphene is governed by the relation $E_F = \hbar v_F \sqrt{\pi n(V_g, P_s, P_f)}$. Below we consider the system at room temperature and so omit the *T*-dependence for the sake of brevity.

Later we'll consider at first the forward sweep in Eq.(A.1, Appendix A), $\frac{dV_g(t)}{dt} > 0$ and start from the case without any dipoles, $P_f = 0, P_s = 0$. However, we assume, that the localized states with the energy $E_T$ and concentration $n_T$ are present at the graphene-substrate interface. For a voltage range $V_g(t) < V_{T1}$, where $V_{T1}$ corresponds to the situation when the occupation of the interface states with the electrons from graphene channel starts, $E_F(V_{T1}) = E_{T1}$, the 2D concentration of electrons in graphene channel is governed by a simple capacitor formula:

$$n(V_g, t) = \frac{\kappa V_g(t)}{4\pi e d}, \qquad (2)$$

where κ is the dielectric permittivity of substrate and *d* is the substrate thickness. The gate voltage $V_{T1}$ leads to the start of the interface states occupation with electrons from the graphene channel ($E_F(V_g(t)) = E_{T1}$). The voltage $V_{T2}$ corresponds to the situation, when all localized interface states are already occupied by electrons from the graphene channel. The voltages are determined by expressions:

$$V_{T1} = \frac{4\pi e d}{\kappa} \frac{E_{T1}^2}{\pi \hbar^2 v_F^2}, \qquad V_{T2} = \frac{4\pi e d}{\kappa} \frac{E_{T2}^2}{\pi \hbar^2 v_F^2} + \frac{4\pi e d n_T}{\kappa}. \qquad (3)$$

In the voltage range $V_{T1} \leq V_g < V_{T2}$, for which the occupation of interface states occur, the concentration in graphene channel remains constant, because additional free carriers, injected from contacts into the graphene channel, are immediately captured by the interface states, and so $n = \frac{E_{T1}^2}{\pi \hbar^2 v_F^2}$. In the voltage range $V_{T2} \leq V_g$ where all the interface states are already occupied by electrons, the free electrons' concentration in the channel is governed by the evident relation

$$n(V_g) = \frac{\kappa V_g(t)}{4\pi e d} - n_T.$$

Let us take into consideration the polarization of dipoles both in the ferroelectric substrate and at the graphene surface. If $P_s(t) \neq 0, P_f(t, T) \neq 0, E_F \leq E_{T1}$, similarly to we get:

$$n(V_g, P_s, P_f) = \frac{\kappa V_g(t)}{4\pi e d} + \frac{P_s(t) + P_f(t, T)}{e} \qquad (4)$$

Below we describe the temperature dependent spontaneous ferroelectric polarization by commonly used expression [32]:

$$P_f(t,T) = P_f(T)\tanh(s_f(V_g(t) - V_c)), \quad (5)$$

where $V_c = E_c d$ is a coercive voltage equal to the product of coercive field and substrate thickness $d$, $s_f$ is a fitting parameter, correspondent to the "sharpness" of ferroelectric switching. The spontaneous polarization of the dipoles absorbed by the graphene surface can be represented as:

$$P_s(t) = P_s \frac{1 - \tanh(s_s(V_g(t) - V_s))}{2} \quad (6)$$

where $V_s$ is some critical voltage, that means that the increase of gate voltage up to some critical value (dependent on the system geometry) finally destroys the polarization ; $s_s$ is a fitting parameter reflecting the "sharpness" of dipoles switching.

Taking into account these two types of polarization, the evident form of Eq. (3)-(4) is listed in Appendix A of Suppl.Mat . These equations are valid for the case, when a lifetime of the carriers trapped by interface states is much greater than the switching time. The validity of this approximation for graphene on PZT substrate was demonstrated experimentally in [33]. Also Eqs. (3)-(4) are valid both for the forward sweep with a sweeping rate $\frac{dV_g}{dt} > 0$, and for the backward one with the negative rate $\frac{dV_g}{dt} < 0$.

For the backward sweep $P_s$ and $P_f$ can be obviously presented as $P_f(t,T) = P_f(T)\tanh(s_f(V_g(t) + V_c))$ and $P_s(t) = P_s\left[1 - \exp\left(-\frac{t(V_s) - t}{\tau}\right)\right]$, where $t(V_s)$ is the moment of time, corresponding to the complete suppression of $P_s$ by the critical gate voltage, $\tau$ is the dipoles' relaxation time that can be rather long of several seconds order, because the absorbed dipoles are obviously non-Langevin ones due to the chemical bonding to graphene surface . Hereinafter we assume that the polarization of the absorbed dipoles starts its renovation immediately after the switching to the backward sweep.

**Results and discussion**

***A. Surface dipoles.*** At first let us consider the case of dielectric substrate without any traps at the graphene-dielectric interface. However, there can be dipoles absorbed by graphene surface (e.g. water molecules, studied experimentally ,[15 - 18]). Therefore, for this case $P_s \neq 0, P_f = 0, n_T = 0$. We also assume, that these dipoles shift the electro-neutrality point into the positive range of $V_g$, as the water molecules do. This means the conductivity of graphene channel is determined by holes that at zero gate voltage.

For this case the carriers concentration in graphene channel is determined both by gate doping and by the absorbed dipoles polarization, which in turn depends on the external electric field, caused by the gate voltage. The carriers concentration is as large as $n_{Ps} = P_s(t)/e \approx 6.23 \times 10^{13}$ cm$^{-2}$ for the polarization value $P_s = 0.1$ C/m$^2$. Therefore it exceeds in two orders of magnitude the maximal concentration $n_{GD} = \kappa V_g/4\pi e d \approx 4.42 \times 10^{11}$ cm$^{-2}$ caused by the gate doping for the range $-8$V $< V_g < +8$V. For the estimation we used the most common parameters of SiO$_2$ substrate $\kappa = 3.9$ and $d = 300$ nm. Therefore the conductivity of the graphene channel would be governed by time-dependent polarization of the dipoles.

**Figure 2** shows the gate voltage dependence of carrier concentration in the graphene channel on SiO$_2$ substrate calculated from Eqs. (4), (6), (A.1) and (A.18) from Suppl. Mat. for different times of surface dipoles relaxation τ and the same switching time. In the initial moment of time (at $V_g = 0$) graphene's channel conductivity is determined by holes, as it was noted before. However, the electric field, caused by the gate potential, destroys the dipoles polarization at some critical value of $V_g$. Note, that the dependences in Figure 2 are presented after the first complete destruction of the surface dipoles polarization; solid lines correspond to the second forward sweep, i.e. polarization disappearance after the first cycle of its renewal. The polarization will start to renew at the backward sweep. The value of polarization that recovers within a switching period is different for different relaxation times τ., but it is clear from the figure, that the polarization renews almost completely within a switching period for a short relaxation time. Therefore the concentration of 2D carriers in graphene is determined by the polarization, and a pronounced hysteresis loop, although asymmetrical one, is characteristic for $n(V_g)$ dependence.

In contrast, the polarization cannot be renewed after being destroyed for the first time for long enough relaxation times, since the 2D carriers concentration is determined by gate doping Eq.(2), and it is, in orders of magnitude smaller according to the previous estimation. Mention, that a direction of such hysteresis loop would be opposite to the so-called direct one, created by ferroelectric dipoles ("anti-hysteresis").

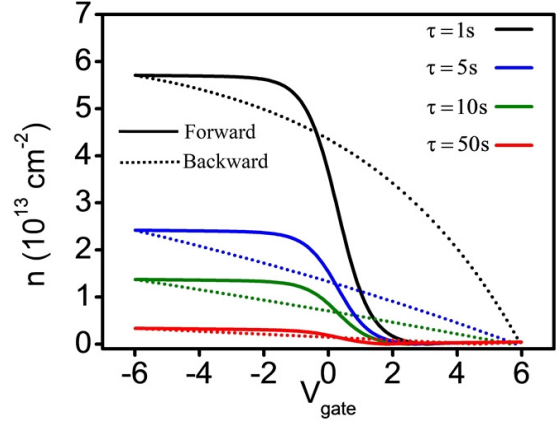

**Figure 2.** The dependence of carrier concentration in the graphene channel on $SiO_2$ substrate vs. the gate voltage calculated for different times $\tau$ of surface dipoles relaxation. Forward and backward sweeping dependences are presented by solid and dotted curves respectively. The curves, calculated for different dipoles' relaxation times $\tau=1, 5, 10$ and $50$ s, are shown by different colours. Parameters, used in calculations are $\kappa=3.9$, $d = 300$ nm, $V_s = 0.3$ V, $dV_g/dt = +5$ V/s for the forward sweep and $dV_g/dt = -5$ V/s for the backward one, $V_g^{max} = \pm 6$ V, $P_s = 0.1$ C/m$^2$ and $s_s = 1$.

**Figure 3** presents the dependence of 2D concentration in graphene channel vs. the gate voltage as a function of the gate voltage switching range. The figure 3 shows that the hysteresis region increases under the gate voltage increase. Considering the direct sweeping (at $dV_g/dt > 0$), it is seen that the disappearance of surface dipoles' polarization occur for all ranges of gate voltages at fixed electric field (see solid curves and Eqs. (4), (6)). After switching the gate voltage in the reverse direction, the dipoles' polarization on the graphene surface immediately starts to recover (see dotted lines and Eq. (A.18)). Therefore, it becomes clear that the value of the polarization that has time to recover will be different for different gate voltage ranges at the same sweeping rate $dV_g/dt$, because the time required to reach different voltage values (-2V, -4V, -6V, -8V) will be different and corresponding polarization value that will have time to recover will be bigger for a higher gate voltage $V_g$.

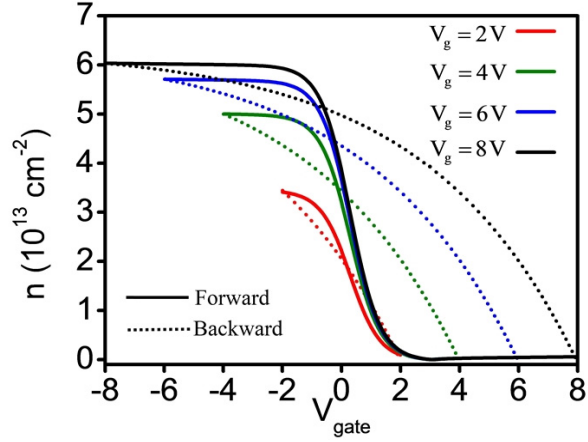

**Figure 3.** The dependence of carrier concentration in the graphene channel on SiO$_2$ substrate vs. the gate voltage calculated for different switching ranges for the gate voltage. Forward and backward sweeping dependences are presented by solid and dotted curves respectively. The curves, calculated for different sweeping range $V_g^{max}$ = ± 2V, ± 4V, ± 6V and ± 8V, are shown by different colours. Parameters used in calculations are κ=3.9, $d$ = 300 nm, $V_s$ = 0.3 V, $\tau$ = 5 s, $dV_g/dt = \pm 1$ V/s; $P_s$ = 0.1 C/m$^2$ and $s_s$ = 1.

**B. Perfect ferroelectric substrate without interface states between the channel and the substrate.** As the next step let us consider the graphene channel on an ideally perfect Pb(Zr$_x$Ti$_{1-x}$)O$_3$ (PZT) substrate without interface states between the channel and the substrate. PZT is widely used as substrate for GFETs ' ' , because its permittivity varies in a broad range depending on the chemical composition $x$, and can be very high near the morphotropic phase boundary at $x$ = 0.52.

It is crucial for our consideration, that PZT response is not a hysteretic one for the electric fields smaller than some coercive value, $E_c$, for which it can be presented by a linear dependence $P(V_g)$ (see works ' and refs. therein). Thus below we will operate within a relatively narrow gate voltage switching range from − 8V to + 8V for which the linear response approximation $P(V_g)$ is valid. Therefore the further analysis should be similar to the one considered above, but the ferroelectric substrate permittivity value is much higher than unity. For $\kappa = 400$ the carrier concentration $n(V_g)$, caused by the gate doping, would be of the same order as the $n(V_g)$ caused by the "doping by absorbed surface dipoles" with $P_s$ = 0.1 C/m$^2$. The maximal value of $n(V_g)$ for the switching range $V_g^{sw} = \pm 8\ V$ can be as large as $n_{GD} = \kappa V_g/4\pi e d \approx 5.89\times 10^{13}\ cm^{-2}$, while the doping by the surface dipoles can be $n_{Ps} \approx 6.23\times 10^{13}\ cm^{-2}$. However, the gate doping creates electron and hole conductivity that depends on the gate voltage polarity, while the doping by surface dipoles for the case of the absorbed water molecules creates the hole conductivity only. Therefore the total concentration of 2D carriers would be determined by combination of the "gate" and "surface dipoles" doping according to Eq.(A.12), (A.14) and (A.18) in Ref..

**Figure 4** presents the carrier concentration in graphene calculated for different dipoles' relaxation times $\tau$. Similarly to the case of $SiO_2$ substrate, the graphene channel on PZT substrate possess the hole conductivity at $V_g = 0$. With a shift of the gate voltage into a positive range the holes concentration decreases and finally reaches the electro-neutrality point ($V_{NP1}$= 1.76 V for $\tau$ = 1 s). In this point the graphene valence band is completely occupied by electrons, and the conduction band is empty. With a further increase of the gate voltage the graphene channel conductivity switches to an electron one, and at the switching point $V_g = 6V$ the electrons concentration is $n = 4.5 \times 10^{13} \, cm^{-2}$.

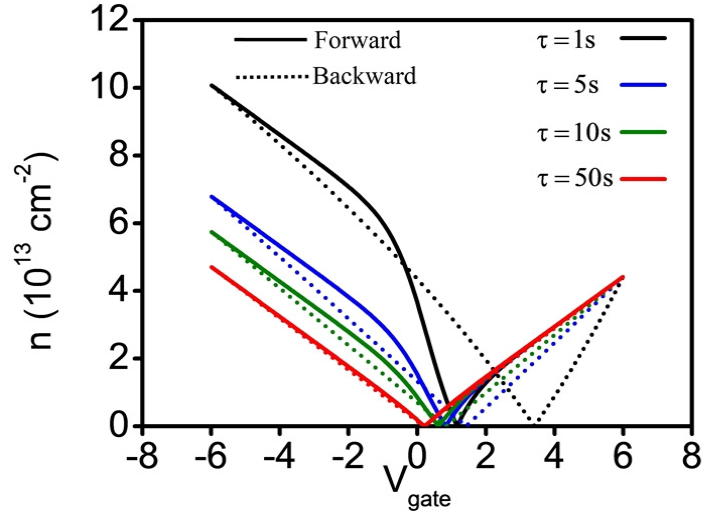

**Figure 4.** The dependence of carrier concentration in the graphene channel on PZT substrate vs. the gate voltage calculated for different relaxation times of surface dipoles. Forward and backward sweeping dependences are presented by solid and dotted curves respectively. The curves, calculated for different dipoles' relaxation times $\tau$=1, 5, 10 and 50 s, are shown by different colours. Parameters used in calculations are $V_s = 0.3$ V, $dV_g/dt = \pm 5$ V/s, $V_g^{max} = \pm 6$ V, $P_s = 0.1$ C/m$^2$, $s_s = 1$, $d = 300$ nm and $\kappa = 400$.

For the backward sweep the electro-neutrality point would occur at higher voltages for a short relaxation time of surface dipoles ($V_{NP2}$ = 3.5 V for $\tau$ = 1 s) than for the forward one ($V_{NP1}$= 1.76 V for $\tau$ = 1 s). With a further change of the gate voltage up to $V_g = -6V$, the conductivity of the graphene channel will be determined by holes with concentration $n \approx 10^{14} \, cm^{-2}$. With the next forward and backward sweeps the hysteresis loop would be reproducible under the same other conditions. The direction of the hysteresis loop, presented in Figs.2-4, is opposite to one another, determined by the ferroelectric dipoles polarization, namely the electro-neutrality point for the backward sweep NP2 shifts to the right from NP1. Under long relaxation times $\tau \geq 50\,s$ the surface polarization cannot restore and corresponding concentration dependence on the gate voltage is described by a linear dependence Eq.(2).

**Figure 5** presents the dependence of carriers' concentration in the graphene channel on PZT substrate vs. the gate voltage calculated for its different amplitude $V_g^{max}$. As anticipated these dependences shows anti-hysteresis behavior and they are qualitatively similar to those for graphene on SiO$_2$ substrate.

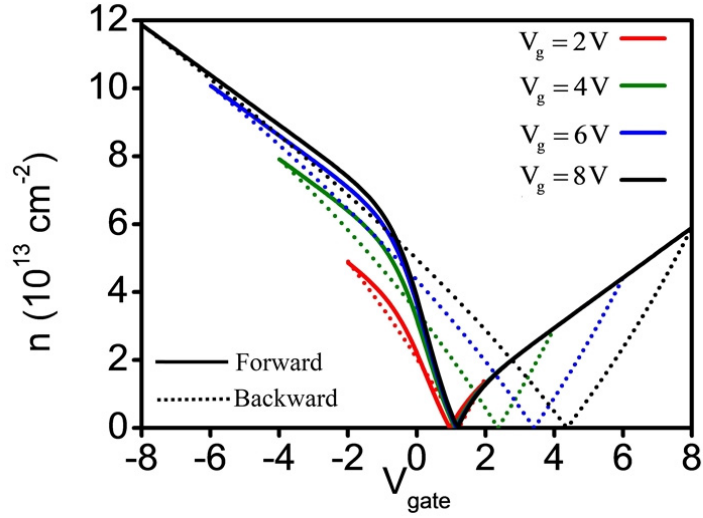

**Figure 5.** The dependence of carrier concentration in the graphene channel on PZT substrate vs. the gate voltage for different amplitudes of gate voltage switching. Forward and backward sweeping dependences are presented by solid and dotted curves respectively. Parameters used in calculations are $d$ = 300 nm, $\kappa$ = 400, $V_s$ = 0.3 V, $\tau$ = 5 s, $dV_g/dt = \pm 1$ V/s, $P_s$ = 0.1 C/m$^2$, $s_s$ = 1, $V_g^{max} = \pm 2$V, $\pm 4$V, $\pm 6$V and $\pm 8$V.

*C. Ideal ferroelectric substrate without surface dipoles.* Let us consider the GFET on an ideal ferroelectric substrate without interface states (as it was done in [34]) and without any surface dipoles, which can be removed by e.g. annealing . In this case $P_s = 0$, $P_f \neq 0$, $n_T = 0$, and so we can take into account the ferroelectric polarization response on the gate voltage only. This response can be described using Landau-Khalatnikov (LK) relaxation equation [35] that is also called time-dependent Ginzburg-Landau equation [36]. It reads

$$\Gamma\frac{dP(t)}{dt}=\alpha(T)P(t)+\beta P^3(t)+\gamma P^5(t)-E(\omega,t) \qquad (7)$$

where the Landau potential expansion coefficients $\alpha = 4.24\times10^5(T-691)$ Jm/C$^2$, $\beta = 4\times0.3614\times10^8$ C$^{-4}\cdot$m$^5$J and $\gamma = 6\times1.859\times10^8$ Jm$^9$C$^{-6}$. All further calculations are performed at room temperature (T=293 K) when the LK relaxation time $\tau_{LK}=\Gamma/|\alpha|$ is very small and has a phonon time order $\sim(10^{-11}-10^{-13})$ s. Note that LK equation can describe the ferroelectric polarization relaxation dynamics in the system "graphene-on-ferroelectric substrate"[37].

Numerical computations according to Eq.(7) enable to obtain the coercive field $E_c(E)$ dependence on applied electric field $E=V_g/d$ created by the gate voltage (**Figure 6**).

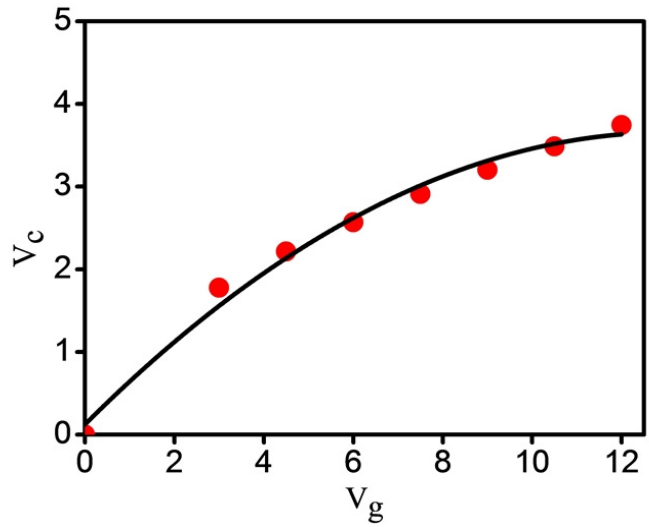

**Figure 6.** Coercive voltage $V_c$ as a function of gate voltage $V_{gate}$ calculated for ferroelectric substrate thickness $d = 300$ nm according to Eq.(7).

**Figure 7** presents the carrier concentration in graphene channel on ideal ferroelectric substrate, $n(V_g)$, as a function of the gate voltage Eqs.(4), (A.13) and (A.17) in Ref.. One can see a pronounced hysteresis in $n(V_g)$ dependence corresponding to the ferroelectric dipoles switching with a change of the gate voltage from the maximal negative value $-V_g^{max}$ to the positive one $+V_g^{max}$. Note, that a distance between electro-neutrality points increases with increasing of the voltage switching range. This hysteresis is caused by the two different stable polarizations of ferroelectric crystal, it can be used for advanced memory cells (see reviews , ). However, the pronounced hysteresis can be observed up to the switching times comparative with the times of ferroelectric domains reversal. This imposes rather strong restrictions on the operating frequency ($10^4$ - $10^6$ Hz) for bulk ferroelectric crystals. However, these times can be several orders of magnitude smaller (about $10^{-9}$s) for thin ferroelectric films of μm thickness and thinner . In our case the nonlinear ferroelectric response will take place at all the gate voltages switching rates within the considered range (0.1 – 50) V/s. This means that the ferroelectric dipoles re-polarization will always follow the change of external field, as well as the hysteresis loop width is constant for a given switching range and does not depend on the sweeping rate. Such a situation was observed experimentally for graphene on PTO/STO substrate with SRO gate .

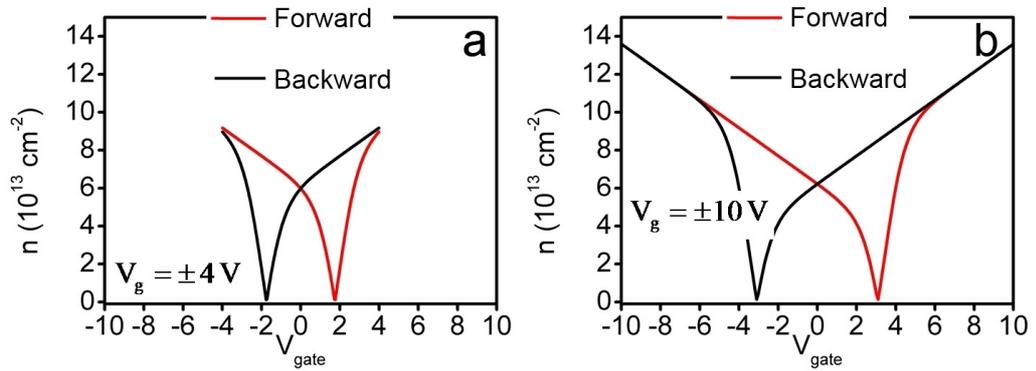

**Figure 7.** The dependence of carrier concentration in the graphene channel on ideal ferroelectric substrate vs. gate voltage calculated for different amplitudes of the gate voltage switching $V_g^{max}$. Forward and backward sweeping dependences are presented by red and black curves respectively. Parameters used in calculations are $\kappa$ =400, $P_f$ = 0.1 C/m$^2$, $d$ = 300 nm, $V_g^{max}$ = ± 4V (a) and $V_g^{max}$ = ± 10V (b).

*D. Gate doping together with doping by ferroelectric substrate and surface dipoles.* Now we can modify the previous case "*C*" by taking into account surface dipoles as well. Therefore we consider the case $P_s \neq 0$, $P_f \neq 0$, $n_T = 0$, so the two rival mechanisms of doping coexist, the first one originates from ferroelectric dipoles and the second one steams from absorbed surface dipoles.

Let us chose the switching range from – 6V to + 6V, for which the ferroelectric response have a form of hysteresis loop with a width $V_c$ = 0.4 V, and take the values of ferroelectric and surface dipoles polarization as $P_s$ = 0.1 C/m$^2$ and $P_f$ = 0.1 C/m$^2$ correspondingly. Since the value of ferroelectric permittivity $\kappa$ = 400 is high enough, the carriers concentration $n_{GD}$, caused by the gate doping for $V_g = \pm 6\,V$, is of the same order of magnitude than the concentration induced by ferroelectric and surface dipoles $n_{P_s}$. In particular, $n_{GD} = \kappa V_g / 4\pi e d \approx 4.42 \times 10^{13}$ cm$^{-2}$ and $n_{P_s} \approx 6.23 \times 10^{13}$ cm$^{-2}$.

**Figure 8** presents the dependence of carriers' concentration in graphene channel on PZT ferroelectric substrate calculated for different rates of gate voltage sweeps (a-b).

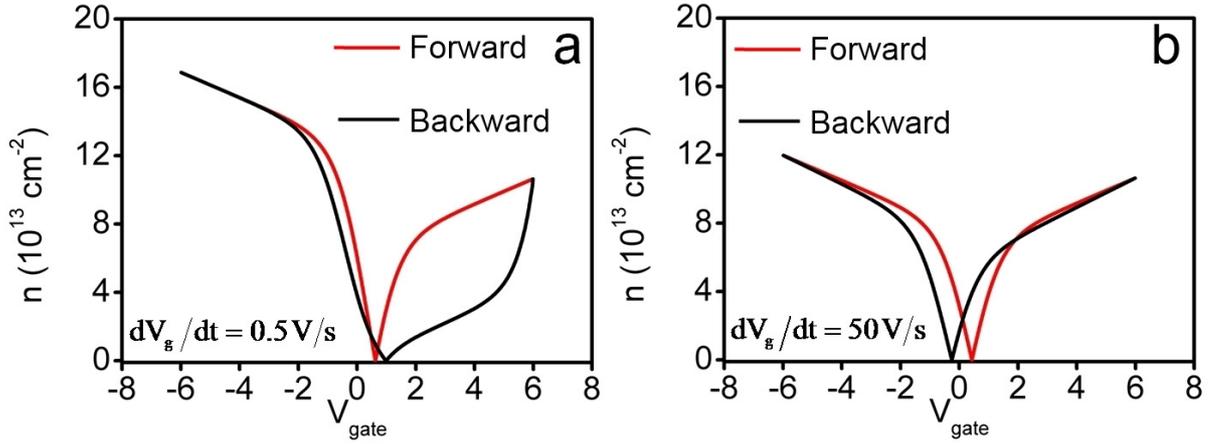

**Figure 8**. The dependence of carriers concentration in the graphene channel with surface absorbed dipoles on real ferroelectric substrate on gate voltage for different rates of gate voltage sweeps. Forward sweep curves are red, the backward ones are black. Parameters, used in the calculations are $d = 300$ nm, $\kappa = 400$, $V_s = 0.2$ V, $V_c = 0.4$ V, $\tau = 1$ s, $V_g^{max} = \pm 6$ V, $P_s = 0.1$ C/m$^2$, $P_f = 0.1$ C/m$^2$, $s_f = 1$, $s_s = 1$, $dV_g/dt = \pm 0.5$ V/s (a), $dV_g/dt = \pm 50$ V/s (b).

As one can see from **Figure 8**, the existence of two rival channels of graphene doping modifies the form of hysteresis loops, presented in previous figures. As it could be expected, the effect of the absorbed surface molecules is essential as slow gate voltage sweeps, and so called anti-hysteresis occurs (Fig.8a). With the increase of $dV_g/dt$ the anti-hysteresis vanishes and finally it transfers into the ferroelectric "direct hysteresis" with the increase of the distance between the electro-neutrality points, up to its final saturation (Fig. 8b).

**Figure 9** presents experimental [38] and theoretical values of electro-neutrality point positions as the function of gate voltage sweeping rate for GFET on PZT ferroelectric substrate. In the experiment a single-layer graphene was fabricated by CVD method on cooper film, and then deposited onto the clean surface of 140 nm PbZr$_{0.2}$Ti$_{0.8}$O$_3$ /60 nm SrRuO$_3$/ SrTiO$_3$ (001) heterostructure. Next the palladium contacts to graphene were launched. As one can see from **Figure 9**, our theory is in a good agreement with experimental data . Small differences between the theoretical and experimental dependences can be caused by the interface states between graphene and ferroelectric, as well as by chemical doping of graphene during its fabrication.

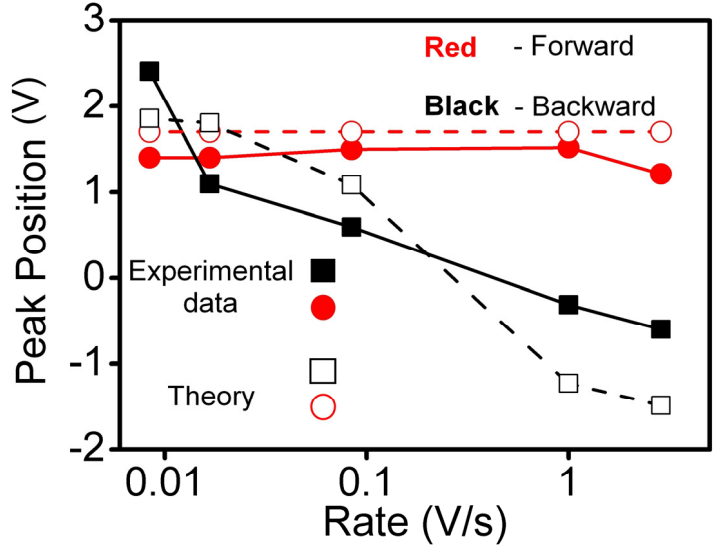

**Figure 9**. Experimental and theoretical values of electro-neutrality point positions as the function of gate voltage sweeping rate for GFET on PZT ferroelectric substrate. Parameters, taken from are: $d$=200 nm, $V_c$ = 2V, $P_f$ = 0,23 C/m$^2$, $V_g^{max}$ ± 5V, for each of the five points $dV_g/dt$ = 0.0084 V/s, 0.017 V/s, 0.084 V/s, 1 V/s, 2.87 V/s, $\tau$ = 20 s. Fitting parameters are $V_s$ = 0, $P_s$ = 0,31 C/m$^2$ and $\kappa$ = 1000.

***E. Gate doping together with doping by ferroelectric substrate and trapping by the interface states.*** Finally we can analyze the more general case, $P_s = 0$, $P_f \neq 0$, $n_T \neq 0$, assuming that the concentration of the interface states is high enough to trap all the electrons injected into the channel due to the gate doping and doping by ferroelectric entire considered $V_g$-range (the validity of this assumption was discussed in ' ). We also assume that there is no saturation of ferroelectric permittivity hysteresis for this range.

Fermi energy in a single layer graphene is governed by expression $E_F = \hbar v_F \sqrt{\pi n}$ . Let us take into consideration the possibility of the carriers trapping from graphene channel by localized states connected with the graphene-surface interface. These states are described by the definite density of states $\rho_T(E)$. We assume that $\rho_T(E)$ is non-zero in some energy interval $E_{T1} < E < E_{T2}$ only, and is equal to zero out of the range.

**Figure 10** demonstrates hysteresis of free carrier concentration $n(V_g)$ in the graphene channel on ferroelectric substrate vs. the gate voltage. The dependences were calculated according to Eqs. (A.12), (A.15a)-(A.16b) in Ref.. using the dependence $P_f(V_g)$ given by Eqs. (A.13) and (A.17) in Ref.., and $V_c(V_g)$ dependence presented in Figure 6.

The form of $n(V_g)$ would be determined in this case by competition of the two rival mechanisms (ferroelectric doping and charge trapping). For the considered case a graphene channel has the holes conductivity at zero gate voltage due to ferroelectric hysteresis. The increase of $V_g$ leads to the decrease of $n(V_g)$ until the electro-neutrality point B would be reached. In this point the graphene valence band is completely occupied by electrons, and the conduction band is empty. With a further increase of the gate voltage the graphene channel conductivity switches to an electron one (B→C) and the concentration $n(V_g)$ would increase until the moment, when the Fermi level would reach the interface states energy ($V_{T1}$, point C), and the electrons trapping by these states starts. In the range $V_{T1} \leq V_g < V_{T2}$ (C→D) where the trapping occurs, the concentration of the free carriers in the graphene channel would be constant, because all the additional carriers, injected into the channel trough the contacts due to the gate and ferroelectric doping, are almost immediately captured by the interface states. After all, i.e. at $V_{T2} \leq V_g(t)$, the interface states would be occupied by electrons and so the dependence $n(V_g)$ would be governed by Eq.(A.16b). Since the lifetime of carriers on the interface states is much longer than the switching time, the backward sweep (D→E→F→G→A) can include both the direct hysteresis (Fig. 10a) and the inverse one ("anti-hysteresis", Fig.10b). Everything is determined by the number of carriers, trapped by interface states, which negative charge screens the electric field in the substrate, that in turn leads to the right-hand-side shift of electroneutrality point along the gate voltage scale, which appears during the backward switching ' , the energy of interface states and the ferroelectric polarization. On the backward sweep graphene would again change its type of conductivity to a hole one (E→F region); in the point F the depletion of the interface states will start, and it would be accomplished in the point G. Finally the hysteresis loop would be closed. For numerical simulations we used the following parameters of the localized states for the electrons trapping and depletion, $E_{T1}$=0,55 eV, $\Delta E = E_{T2} - E_{T1}$= 50 μeV, $E_{T1(hole)}$=0,83 eV and $\Delta E_{(hole)}=E_{T2(hole)} - E_{T1(hole)}$= 50 μeV. The concentration of electrons, trapped by interface states, is $n_T = 6.75 \times 10^{13}$ cm$^{-2}$ (a) and $n_T = 9.16 \times 10^{13}$ cm$^{-2}$ (b).

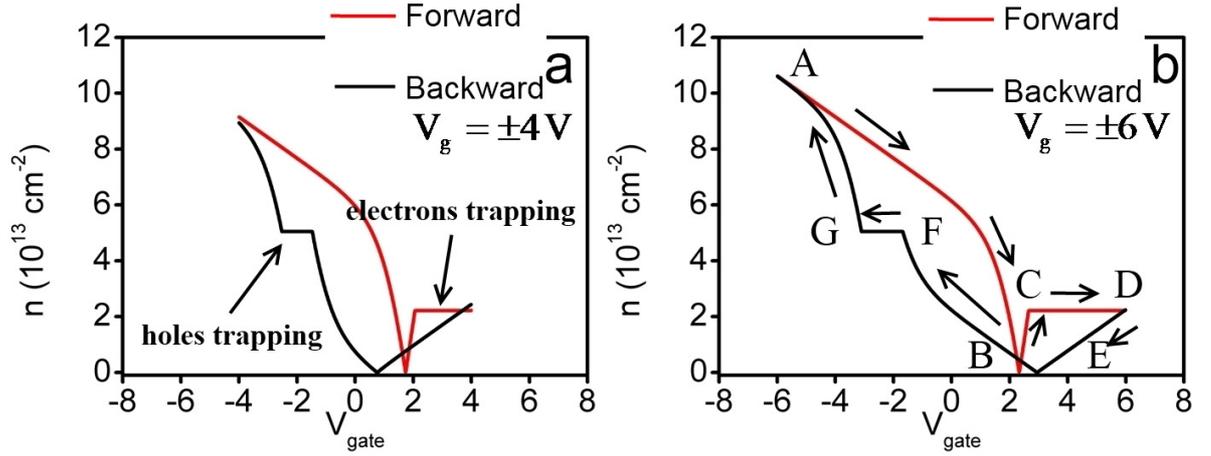

**Figure 10**. Hysteresis dependence of free carrier concentration in graphene channel on ferroelectric substrate vs. the gate voltage. The direction of loop is A→B→C→D→E→F→G→A. The parameters $d$ = 300 nm, $\kappa$=400, $P_f$ = 0,1 C/m$^2$, $V_g^{max} \pm$ =4V (a) and $V_g^{max} \pm$ =6V (b). Forward sweeping dependences are shown by red curves, the backward one by the black ones.

The dependences qualitatively similar to those shown in **Figures 10** were obtained experimentally for the graphene channel placed on hexagonal BN (hBN) on (1-x)[Pb(Mg$_{1/3}$Nb$_{2/3}$)O$_3$]-x[PbTiO$_3$] (PMN-PT) substrate [39]. The transition from the direct hysteresis to inverse one was observed there with the increase of the gate voltage switching range, as it is demonstrated in **Figure 10.**

**Conclusion**
Using a phenomenological approach, we developed a general theory for the analytical description of versatile hysteretic phenomena in GFETs on different substrates. Evolved formalism considers different origin of the existence of the external dipoles on graphene free surface and the localized states at the graphene-surface interface.

We demonstrated that the dipole molecules absorbed by the graphene surface (e.g. polar water molecules) can cause hysteretic form of the graphene conductivity dependence on a gate voltage $n(V_g)$ in GFET on different substrates. It was demonstrated, that the increase of the gate voltage sweeping rate $dV_g/dt$ leads to the complete vanishing of hysteresis both for GFET on SiO$_2$ substrate and on ferroelectric substrate for the gate voltages less than the coercive one, $V_g<V_c$. These results are in qualitative and quantitative agreement with the available experimental data . Also we have shown that the increase of gate voltage rate causes the transition from anti-hysteresis to ferroelectric hysteresis for PZT substrate, that correlates well with the experiment .

Notice that our results have been obtained within a number of essential approximations and simplifications, such as the approximation for spontaneous relaxation of the absorbed dipoles polarization with a single relaxation time τ. More rigorous model should be based on deeper understanding of the nature of dipoles bonding to graphene surface and account for a possible relaxation time spectrum. Also we treated the process of the carriers trapping by surface states as a very swift process independent on the frequency of gate voltage switching. However, it was demonstrated for GFET on $PbTiO_3/SrTiO_3$ substrate in ref., that the process of carriers trapping becomes less effective for the gate voltage switching frequency higher than 1 KHz that corresponds to the sweeping rate $10^4$ V/s. The understanding of the effect is possible on the base of the clear vision of the physical mechanism of trapping and the configuration of its potential.

Since GFETs on ferroelectric substrate are very promising candidates for the non-volatile ultra-fast memory of new generation (see e.g. ' ' ), the theoretical description of the devices operation requires reliable knowledge about the ferroelectric response at low, intermediate and high frequencies, as well as about the processes of carrier trapping at a given frequency range. This knowledge is crucial, because the competition of the abovementioned factors would determine the FRAM operation characteristics. Obtained results are valid for the description of hysteretic phenomena in various realistic GFETs operating at low and intermediate frequency range, may be not directly applicable for the description of ferroelectric response at ultra-high frequencies, priory because the high-frequency response of ferroelectric dipoles requires further experimental studies.

**Conflict of interests.** Authors declare no conflict of interests of any kind.

# Supplementary Materials to
# "Hysteretic phenomena in GFET: general theory and experiment"

Anatolii I. Kurchak[1], Anna N. Morozovska[2], and Maksym V. Strikha[1,3]*

[1] *V.Lashkariov Institute of Semiconductor Physics, National Academy of Sciences of Ukraine,*
*pr. Nauky 41, 03028 Kyiv, Ukraine*
[2] *Institute of Physics, National Academy of Sciences of Ukraine,*
*pr. Nauky 46, 03028 Kyiv, Ukraine*
[3] *Taras Shevchenko Kyiv National University, Radiophysical Faculty*
*pr. Akademika Hlushkova 4g, 03022 Kyiv, Ukraine*

## Appendix A. Calculations details

Saw-like dependence of $V_g$ on time $t$ (Fig.A.1) can be presented for calculations in Fourier series:

$$V_g(t) = \sum_n \frac{8A}{\pi^2}\left((-1)^n \frac{\sin((2n+1)\omega t)}{(2n+1)^2}\right), \qquad (A.1)$$

where $A$ is an amplitude. Later we'll restrict ourselves it numerical calculations by the first 100 addends in Eq.(A.1), which have a sufficient accuracy.

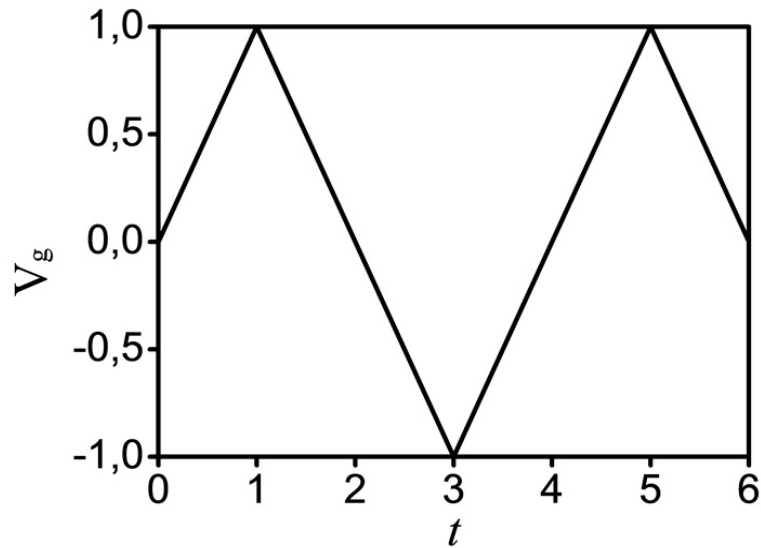

**Figure A1.** Schematic saw-like dependence of gate voltage on time (all in arbitrary units).

---

* corresponding author, e-mail: maksym_strikha@hotmail.com

The resistivity of graphene channel can be presented similarly to [1] as:

$$\rho(V_g(t), P_s(t), P_f(t,T), T) \approx \frac{1}{\sigma(V_g(t), P_s(t), P_f(t,T))} + \frac{1}{\sigma_{intr}(T)} + \frac{1}{\sigma_{min}}, \quad (A.2)$$

where $P_s(t)$ is the polarization of dipoles, absorbed on graphene surface, $P_f(t,T)$ is the temperature-dependent ferroelectric polarization.

In the case when the scattering of ionized impurities by substrate dominate (the most common case for graphene in ambient) the conductivity of 2D graphene channel can be presented as :

$$\sigma(V_g(t), P_s(t), P_f(t,T)) = e n(V_g(t), P_s(t), P_f(t,T)) \mu \quad (A.3)$$

where $n(V_g(t), P_s(t), P_f(t,T))$ is the 2D concentration of carriers per unit area, caused my mixed doping by gate, surface absorbed dipoles and ferroelectric dipoles, $\mu$ is the carriers mobility, which also are the functions of temperature $T$.

The second term in Eq.(A.2) corresponds the intrinsic graphene conductivity, which should be taken into consideration if the Fermi level is close to the electro-neutrality point:

$$\sigma_{intr}(T) = e n_{intr}(T) \mu \quad (A.4)$$

$$n_{intr}(T) = \frac{2(kT)^2}{\pi(\hbar v_F)^2}. \quad (A.5)$$

The third term in Eq.(A.2) corresponds the minimal quantum conductivity, which is to be taken into consideration at low $T$.

$$\sigma_{min} \approx \frac{4e^2}{\hbar}. \quad (A.6)$$

Fermi energy in a single-layer graphene is given by relation [2]:

$$E_F = \hbar v_F \left(\pi n(V_g(t), P_s(t), P_f(t,T))\right)^{1/2} \quad (A.7)$$

At first let's analyze the forward sweep in Eq.(A.1), $\frac{dV_g(t)}{dt} > 0$, for the case without any dipoles, $P_f = 0$, $P_s = 0$. Also we assume, that the localized states with energy $E_T$ and concentration $n_T$ are present at the graphene-substrate interface. For voltage range $V_g(t) < V_{T1}$, where $V_{T1}$ corresponds to the condition $E_F(V_{T1}) = E_{T1}$ (the occupation of the interface states with the electrons from graphene channel starts), the 2D concentration of electrons in graphene channel is governed by a simple capacitor formula [3, 4]:

$$n(V_g(t)) = \frac{\kappa V_g(t)}{4\pi e d} \quad (A.8)$$

where κ is dielectric permittivity and $d$ is the thickness of substrate The gate voltage $V_{T1}$ leads to the start of the interface states occupation with electrons from the graphene channel ($E_F(V_g(t)) = E_{T1}$).

The voltage $V_{T2}$ corresponds to the situation, when all localized interface states are already occupied by electrons from the graphene channel. The voltages [1,]:

$$V_{T1} = \frac{4\pi e d}{\kappa} \frac{E_{T1}^2}{\pi \hbar^2 v_F^2} \quad, \quad V_{T2} = \frac{4\pi e d}{\kappa} \frac{E_{T2}^2}{\pi \hbar^2 v_F^2} + \frac{4\pi e d n_T}{\kappa} \tag{A.9}$$

In the voltage range $V_{T1} \leq V_g < V_{T2}$, where the occupation of interface states occur, the concentration in graphene channel remains constant, because additional free carriers, injected from contacts into the graphene channel, are immediately captured by interface states[4]:

$$n = \frac{E_{T1}^2}{\pi \hbar^2 v_F^2} \tag{A.10}$$

For the range $V_{T2} \leq V_g$, when all the interface states are already occupied by electrons, the free electrons concentration in the channel is governed by obvious relation:

$$n(V_g(t)) = \frac{\kappa V_g(t)}{4\pi e d} - n_T \tag{A.11}$$

Now we'll take into consideration the polarization of dipoles both in the ferroelectric substrate and at the graphene surface. If $P_s(t) \neq 0$, $P_f(t, T) \neq 0$, $E_F \leq E_{T1}$, similarly to [1, 3, 4] we get:

$$n(V_g(t), P_s(t), P_f(t,T)) = \frac{\kappa V_g(t)}{4\pi e d} + \frac{P_s(t) + P_f(t)}{e}. \tag{A.12}$$

We describe the temperature dependent spontaneous polarization of ferroelectric substrate by commonly used expression:

$$P_f(t,T) = P_f(T)\tanh(s_f(V_g(t) - V_c)) \tag{A.13}$$

where $V_c = E_c d$ is coercive voltage, equal to the product of coercive field and ferroelectric thickness $d$, $s_f$ is a fitting parameter, correspondent to the 'sharpness' of ferroelectric switching.

The spontaneous polarization of the dipoles absorbed by graphene surface can be presented as:

$$P_s(t) = P_s \frac{1 - \tanh(s_s(V_g(t) - V_s))}{2} \tag{A.14}$$

where $V_s$ is some critical voltage (the increase of gate voltage to some critical value, dependent on system geometry, finally ruins this polarization ), $s_s$ is a fitting parameter, correspondent to the 'sharpness' of dipoles switching.

Taking into account these two types of polarization, Eq. (A.9-11) can be rewritten as:

$$V_{T1} = \frac{4\pi e d}{\kappa} [\frac{E_{T1}^2}{\pi \hbar^2 v_F^2} - \frac{(P_s(t) + P_f(t,T))}{e}] \quad, \tag{A.15a}$$

$$V_{T2} = \frac{4\pi e d}{\kappa}[\frac{E_{T2}^2}{\pi\hbar^2 v_F^2} + n_T - \frac{(P_s(t) + P_f(t,T))}{e}], \qquad (A.15b)$$

$$n = \frac{E_{T1}^2}{\pi\hbar^2 v_F^2} - \frac{P_s(t) + P_f(t,T)}{e}, \qquad (A.16a)$$

$$n(V_g) = \frac{\kappa V_g}{4\pi e d} + \frac{P_s(t) + P_f(t,T)}{e} - n_T \qquad (A.16b)$$

Note, that Eqs.(A.15)-(A.16) are valid for the case, when a life time of carriers, trapped by interface states, is much greater than the switching time. The validity of the approximation for graphene on PZT substrate was demonstrated experimentally in [5].

Eqs. (A.15)-(A.16) are valid both for the forward sweep ($dV_g(t)/dt > 0$), and for the backward one ($dV_g(t)/dt > 0$). However, for the backward sweep $P_s$ and $P_f$ should be obviously presented as:

$$P_f(t,T) = P_f(T)\tanh(s_f(V_g(t) + V_c)) \qquad (A.17)$$

$$P_s(t) = P_s\left[1 - exp\left(-\frac{t(V_s) - t}{\tau}\right)\right] \qquad (A.18)$$

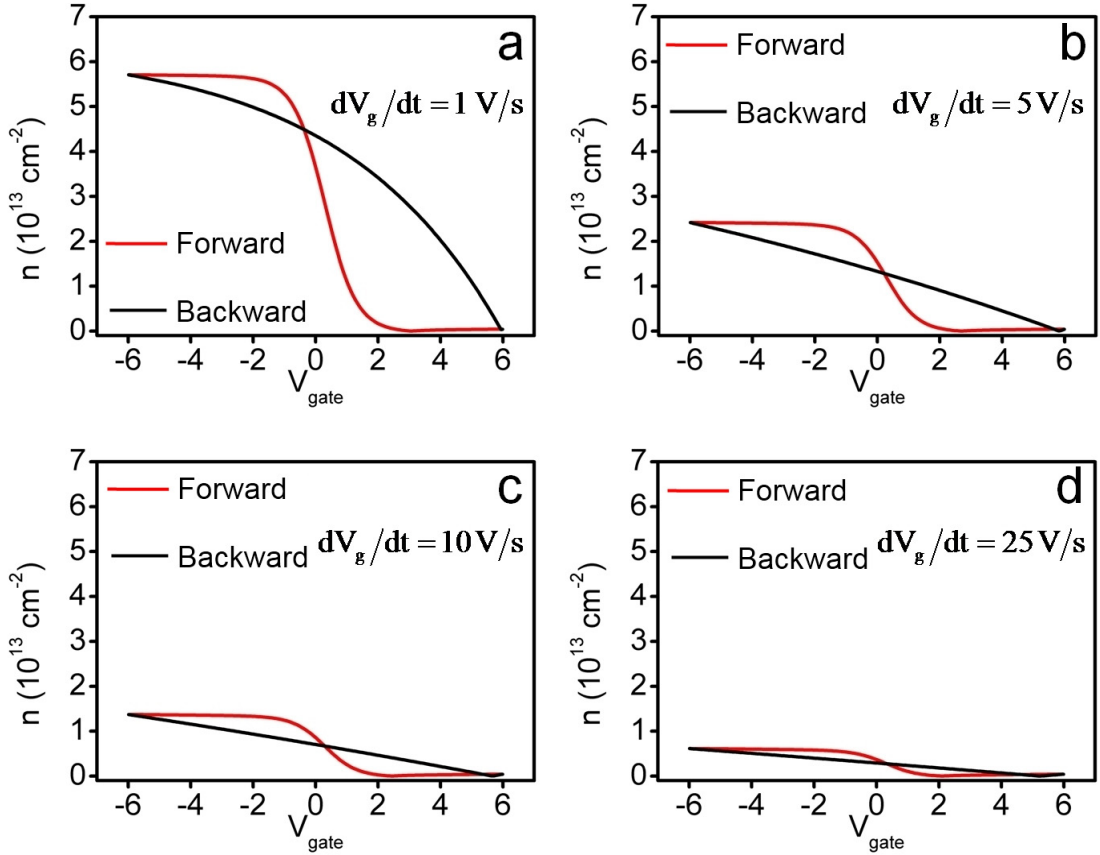

**Figure A2.** The dependence of carrier concentration in the graphene channel on SiO$_2$ substrate on gate voltage for different sweeping rates of the gate voltage. Forward sweep ($dV_g(t)/dt > 0$) is shown by red curves, backward one ($dV_g(t)/dt < 0$) – by the black ones. Parameters, used for calculations: κ=3.9; $V_s$ = 0.3 V; τ = 5 s; $V_g$ = 6 V; $P_s$ = 0.1 C/m²; $s_s$ = 1; a) $dV/dt$ = 1 V/s; b) $dV/dt$ = 5 V/s; c) $dV/dt$ = 10 V/s; d) $dV/dt$ = 25 V/s.

**Figure A.2** presents the dependence of 2D concentration in graphene channel vs. the gate voltage as a function of the gate voltage switching rate. Under slow sweeping rates the surface dipoles polarization renews, under fast ones it vanishes completely. This corresponds qualitatively the experimental data [6] for the graphene channel on $SiO_2$ substrate with the absorbed water molecules on it's surface. At that the increase of $dV_g/dt$ leads to narrowing and later on to vanishing of hysteresis loop in $\sigma(V_g)$ dependence of the graphene FET with Si gate.

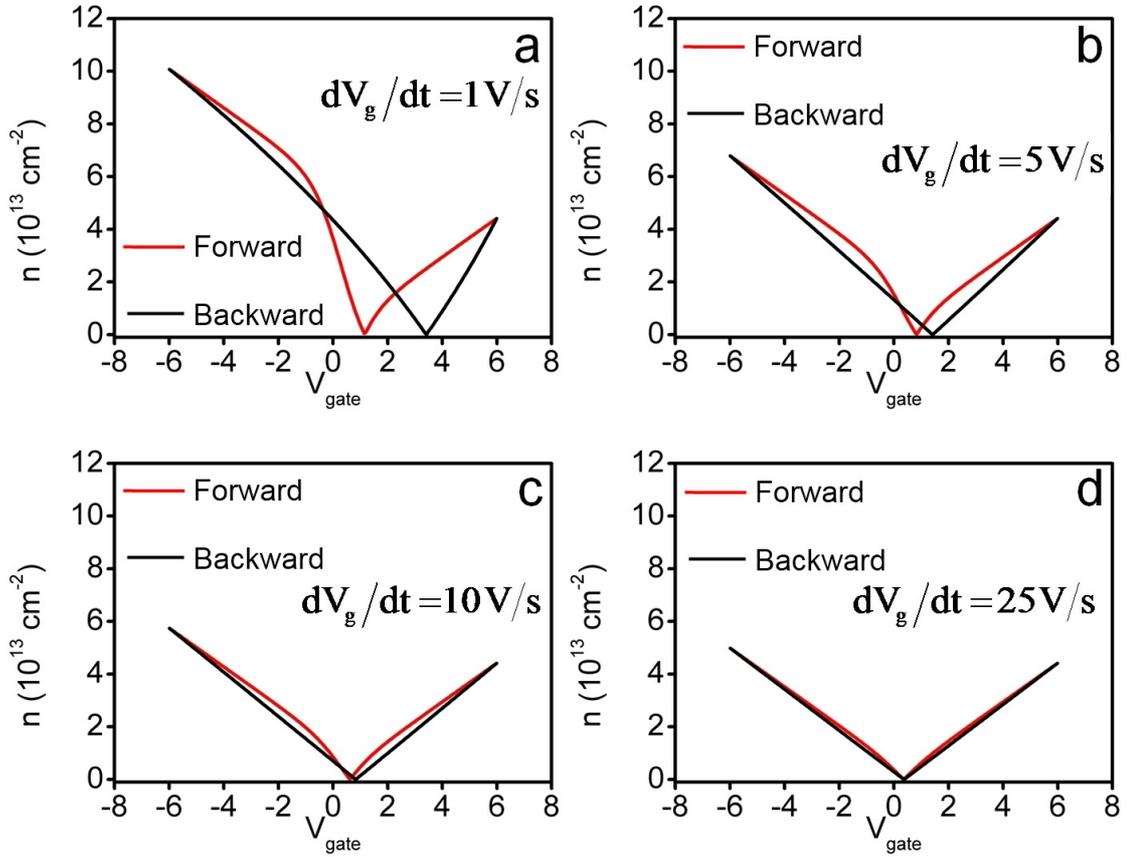

**Figure A3.** The dependence of carrier concentration in the graphene channel on PZT substrate on gate voltage for different rates of gate voltage sweeps. Forward sweep ($dV_g(t)/dt > 0$) is shown by red curves, backward one ($dV_g(t)/dt < 0$) – by the black ones. Parameters, used for calculations: κ = 400; $V_s$ = 0.3 V; τ = 5 s; $V_g$ = 6 V; $P_s$ = 0.1 C/m²; $s_s$ = 1; a) $dV/dt$ = 1 V/s; b) $dV/dt$ = 5 V/s; c) $dV/dt$ = 10 V/s; d) $dV/dt$ = 25 V/s.

**Figure A.3** presents the dependence of 2D concentration in graphene channel vs. the gate voltage as a function of the gate voltage switching rate. Under slow sweeping rates the surface dipoles polarization renews, under fast ones it vanishes completely.